\documentclass[pra,twocolumn,superscriptaddress,showpacs,floatfix,longbibliography]{revtex4-2}
\usepackage{mathrsfs,braket}
\usepackage{amssymb, amsbsy, amsmath, latexsym, dsfont, array, layout,
graphicx,mathrsfs,color,ulem,bm}
\usepackage[colorlinks=true,citecolor=blue,urlcolor=blue]{hyperref}

\begin{document}

\title{Tuning excitation transport in a dissipative Rydberg ring}

\author{Yiwen Han}
\affiliation{CAS Key Laboratory of Quantum Information, University of Science and Technology of China, Hefei 230026, China}
\author{Wei Yi}
\email{wyiz@ustc.edu.cn}
\affiliation{CAS Key Laboratory of Quantum Information, University of Science and Technology of China, Hefei 230026, China}
\affiliation{CAS Center For Excellence in Quantum Information and Quantum Physics, Hefei 230026, China}

\begin{abstract}
We demonstrate the flexible tunability of excitation transport in Rydberg atoms, under the interplay of controlled dissipation and interaction-induced synthetic flux. Considering a minimum four-site setup---a triangular configuration with an additional output site---we study the transport of a single excitation, injected into a vertex of the triangle, through the structure. While the long-range dipole-dipole interactions between the Rydberg atoms lead to geometry-dependent Peierls phases in the hopping amplitudes of excitations, we further introduce on-site dissipation to a vertex of the triangle. As a result, both the chirality and destination of the transport can be manipulated through the flux and dissipation. In particular, we illustrate a parameter regime where our Rydberg-ring structure may serve as a switch for transporting the injected excitation through to the output site. The underlying mechanism is then analyzed by studying the chiral trajectory of the excitation and the time-dependent dissipation.
\end{abstract}

\maketitle

\section{Introduction}

In recent years, Rydberg atoms have proved to be an ideal candidate for exploring intriguing synthetic quantum matter~\cite{reviewRdB-1,reviewRdB-2,reviewRdB-3,reviewRdB-4}. The highly flexible geometry facilitated by individual optical-tweezer trapping~\cite{Ashkin-86,Ebadi-21,Morgado-21,Scholl-21}, together with the long-range dipole-dipole interactions~\cite{Weber-17,Browaeys-16}, offer abundant control degrees of freedom for simulating a variety of many-body quantum models and strongly correlated phenomena, including quantum spin models~\cite{Leseleuc-18,Bernien-17,Schauss-18,Barredo-15,Signoles-21,Steinert-23}, the topological order~\cite{Leseleuc-19,Verresen-21,Weber-22}, dynamic gauge fields~\cite{Zhang-18,Celi-20,Notarnicola-20,Surace-20,Cuadra-22}, and quantum spin liquids~\cite{Giudici-22,Semeghini-21,Cheng-23,Kalinowski-23,Ohler-23}. Particularly, under dipolar exchange interactions, Rydberg atoms can be modeled as few-level systems in the $nS$ and $nP$ manifolds, with excitations in the $nP$ states hopping between different atoms that are often spatially separated~\cite{Lienhard-20}. When more than one excited states per atom are involved, the dipole interaction couples the internal degrees of freedom with the center-of-mass motion of an excitation, leading to synthetic gauge fields (in the form of either spin-orbit coupling or the Peierls phase) that are responsible for the topological bands~\cite{Zhao-23,F-23,Yang-22} and chiral motion~\cite{Lienhard-20,Wu-22,Li-22} of the Rydberg excitations.

While the Rydberg excitations are known for their remarkable life time, they are also amenable to engineered dissipation, for instance, by coupling the $nP$ state to some short-lived intermediate states. This would open up fresh possibilities for emulating quasiparticles in solid-state materials, where the quasiparticle excitations naturally acquire a finite life time due to their interactions with the background environment. The introduction of dissipation also makes connection with the ongoing study of non-Hermitian physics~\cite{reviewNH-1,Shen-23}, where the interplay of synthetic gauge field and dissipation often gives rise to intriguing dynamics.

In this work, we study the transport of a single Rdyberg excitation through a triangular ring structure, where the dipole-induced synthetic flux and dissipation offer flexible control over the transport dynamics. In a recent experiment, the chiral motion of a single excitation was observed in a minimum triangular setup of Rydberg atoms~\cite{Lienhard-20}.
The chiral motion relies on the spin-orbit coupling intrinsic to the dipole interactions, as well as external magnetic and electric fields that break the time-reversal symmetry and the degeneracy of the $nP$ states. Building upon the experiment, we consider a four-site configuration, where a triangular ring is attached to an output site. Under the long-range dipolar exchange interaction, the hopping amplitude of the excitation between any two sites acquires a Peierls phase that is sensitive to the relative position of the sites~\cite{Lienhard-20}. Since the four-site configuration we consider contains three triangles, each threaded by a geometry-dependent synthetic magnetic flux, the transport dynamics of the excitation is a consequence of the competition between the various fluxes. Specifically, we consider the dynamics of a single excitation injected into a vertex of the triangle, transported through the triangular ring structure to the output site.
In the absence of dissipation, the flux-dependent chiral motion in the simple triangular case becomes more complicated, and shows a counter-clockwise (CCW) tendency, regardless of the flux parameters.
We then introduce a local on-site dissipation to one of the sites (other than the output site), and show  that, through the simultaneous tuning of the flux and dissipation, the chiral motion of the excitation can be restricted to within different triangular sites, such that the transport to the output site can be switched on demand. We characterize the transport dynamics through the chiral trajectory of the excitation, as well as the total loss rate of the excitation, which offer a transparent picture for the tunable transport dynamics. Our findings suggest the interplay between the interaction-induced synthetic gauge field and dissipation provides unique opportunities for quantum control and simulation in Rydberg atoms.

Our work is organized as follows. In Sec.~II, we introduce the four-site configuration as well as the resulting model Hamiltonian. We present the main results in Sec.~III and Sec.~IV, where we study the transport dynamics either in the presence or absence of dissipation. We also show the ideal parameter regime to control the chiral motion and transport dynamics in Sec.~IV. In Sec.~V, we analyze the transport dynamics using chiral trajectory and decay rate, with increasing dissipation. We summarize in Sec.~VI.

%
%

\section{Configuration and Model}

\begin{figure}[tbp]
\begin{centering}
\includegraphics[width=0.48\textwidth]{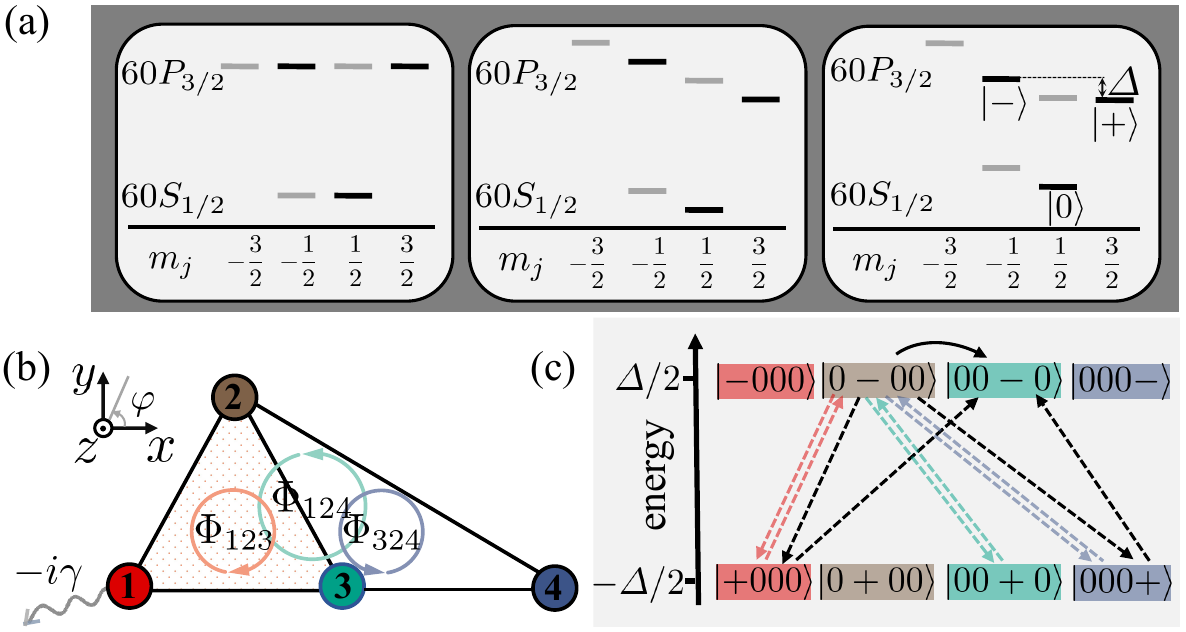}
\par\end{centering}
\caption{\label{fig:1} Synthetic flux induced by dipolar exchange interaction.
(a) Schematic Zeeman structure in the relevant Rydberg manifolds, $60S_{1/2}$
and $60P_{3/2}$. The sub-figures from left to right respectively illustrate the original level scheme considering the quantum-defect effect, the level scheme under the Zeeman shifts, and that affected  the dc stark effect additionally.
(b) Four atoms are confined in the $x$-$y$ plane. Excitations are injected into atom $2$, and an on-site dissipation is introduced to atom $1$. The interaction-induced synthetic fluxes $\Phi_{ijk}$ are defined in the main text.
(c) The second-order processes
described by the colored dashed lines are included in the self-energy
of the $\left|0-00\right\rangle $ state. The transition from
$\left|0-00\right\rangle $ to $\left|00-0\right\rangle $ consists of two processes: a direct
hopping with an amplitude $t_{-}$ (black solid arrow), and
the virtual hoppings through $\left|+000\right\rangle $ or $\left|000+\right\rangle $
(black dashed lines). Here the notation $|\pm 000\rangle$ indicates the first atom in an excited $|\pm\rangle$ state.}
\end{figure}

Our system comprises four $^{87}\text{Rb}$ atoms confined in optical
tweezers. For each atom, the relevant three Rydberg states in the $60S_{1/2}$ and $60P_{3/2}$ manifolds are marked by
bold black lines in Fig.~\ref{fig:1}(a). Following Ref.~\cite{Lienhard-20}, we label the states as
$\left|+\right\rangle =\left|60P_{3/2},m_{j}=3/2\right\rangle $,
$\left|-\right\rangle =\left|60P_{3/2},m_{j}=-1/2\right\rangle $, and $\left|0\right\rangle =\left|60S_{1/2},m_{j}=1/2\right\rangle $,
where states $\left|\pm\right\rangle$
correspond to the two internal states of the Rydberg excitation, and $|0\rangle$ indicates absence of excitation.
We assume the states $|\pm\rangle$ are isolated from others in $60P_{3/2}$ manifold by
static magnetic and electric fields (in the $z$ direction), and the degeneracy of the two states are broken, with an energy difference $\varDelta=E_{-}-E_{+}$.

We consider a four-site configuration in the $x$-$y$ plane, as illustrated in Fig.~\ref{fig:1}(b). Atoms on sites $1$, $2$ and $3$ form an equilateral triangle (each side of length $r_0$), while a fourth atom is located on the output site (labeled $4$). The distance between atoms $3$ and $4$ is also $r_0$.


The excitation transport between any two sites $i$ and $j$ is driven by the dipole-dipole interaction~\cite{Peter-15}
\begin{align}
\hat{V}_{ij}&=\text{\ensuremath{\frac{1}{4\pi\epsilon_{0}r_{ij}^{3}}}}\left[\hat{d}_{i}^{z}\hat{d}_{j}^{z} +\frac{1}{2}\left(\hat{d}_{i}^{+}\hat{d}_{j}^{-}+\hat{d}_{i}^{-}\hat{d}_{j}^{+}\right)\right.\nonumber\\ &\left.-\frac{3}{2}\left(\hat{d}_{i}^{+}\hat{d}_{j}^{+}e^{-i2\ensuremath{\varphi}_{ij}}+\hat{d}_{i}^{-}\hat{d}_{j}^{-}e^{i2\ensuremath{\varphi}_{ij}}\right)\right],\label{eq:1}
\end{align}
where the relative position of the atoms $r_{ij}=\left|\mathbf{r}_{j}-\mathbf{r}_{i}\right|$,
and the position-dependent phase $\varphi_{ij}=\arg\left(\mathbf{r}_{j}-\mathbf{r}_{i}\right)$.
Components of the dipole operator are given by
$\hat{d}_{i}^{x}$, $\hat{d}_{i}^{y}$,
$\hat{d}_{i}^{z}$, and $\hat{d}_{i}^{\pm}:=\mp\left(\hat{d}_{i}^{x}\pm i\hat{d}_{i}^{y}\right)/\sqrt{2}$.
Of particular interest here are the last terms in the brackets, namely $\hat{d}_{i}^{+}\hat{d}_{j}^{+}e^{-i2\ensuremath{\varphi}_{ij}}$ and its conjugate,
which formally represent the dipole-induced spin-orbit coupling where the internal-state degrees of freedom and the center-of-mass motion (represented by the phase) are coupled.

We then introduce light-induced dissipation to the excited $\left|\pm \right\rangle$ states of atom $1$, characterized by a decay rate $\gamma$. This can be engineered by resonantly coupling these long-lived excited states to short-lived states, for instance, in the $6S$ manifold.
Dynamics of atoms in the $60S_{1/2}$ and $60P_{3/2}$ manifolds are hence driven by an effective non-Hermitian Hamiltonian denoted as $H_{\text{full}}$, which resides in the single-excitation subspace spanned by the states $\left\{ \left|\pm 000\right\rangle ,\left|0\pm 00\right\rangle , \left|00\pm 0\right\rangle ,\left|000\pm\right\rangle \right\} $ (see Appendix).
For the dynamic transport, we initialize the system in the state $\left|0-00\right\rangle $, where atom $2$ is in the $\left|-\right\rangle $ excited state, while the other
atoms are in the ground $\left|0\right\rangle $ state. Assuming a large energy difference $\varDelta$ between the $|+\rangle$ and $|-\rangle$ excitations, the system should primarily remain in the subspace with the basis states $\left\{ \left|-000\right\rangle ,\left|0-00\right\rangle ,\left|00-0\right\rangle ,\left|000-\right\rangle \right\} $.
We then derive the effective Hamiltonian by eliminating the $|+\rangle$ excitations,
which, in a second-quantized form, reads
\begin{equation}
H_{\text{eff}}=\sum_{i}\varepsilon_{i}a_{i}^{\dagger}a_{i}+\sum_{ij}t_{ij}e^{i\phi_{ij}}a_{i}^{\dagger}a_{j}.
\end{equation}
Here $a^\dag_i$ ($a_i$) creates (annihilates) an excitation on site $i$ in the state $|-\rangle$.
The derivation and expressions for the self-energy terms $\varepsilon_i$ and the hopping terms $t_{ij}e^{i\phi_{ij}}$ are given in the Appendix.

Importantly, the hopping amplitudes of the excitations carry Peierls phases $\phi_{ij}$, which contribute to the synthetic fluxes $\Phi_{ijk}$ shown in Fig.~\ref{fig:1}(b).
More explicitly, we have the expression $\Phi_{ijk}\equiv\phi_{ij}+\phi_{jk}+\phi_{ki}$, defined within the range of $\left(-\pi,\pi\right]$.
Since we take a second-order perturbation and eliminate the $|+\rangle$ excitations,
the phase $\phi_{ij}$ includes contributions from both the direct hopping of the $|-\rangle$ excitation from site $i$ to site $j$ (that is, $|-\rangle_i\rightarrow |-\rangle_j$), and those mediated by a $|+\rangle$ excitation on a third site $k$ (that is, $|-\rangle_i\rightarrow |+\rangle_k \rightarrow |-\rangle_j$). For an explicit example regarding the transition from
$\left|0-00\right\rangle $ to $\left|00-0\right\rangle $, see Fig.~\ref{fig:1}(c).
As a result, $\phi_{14}\neq \phi_{13}+\phi_{34}$ as they represent distinct physical processes. Likewise, $\Phi_{124}\neq \Phi_{123}+\Phi_{324}$, unlike the Raman-induced Peierls phases for neutral atoms.

\begin{figure}[tbp]
\includegraphics[width=0.48\textwidth]{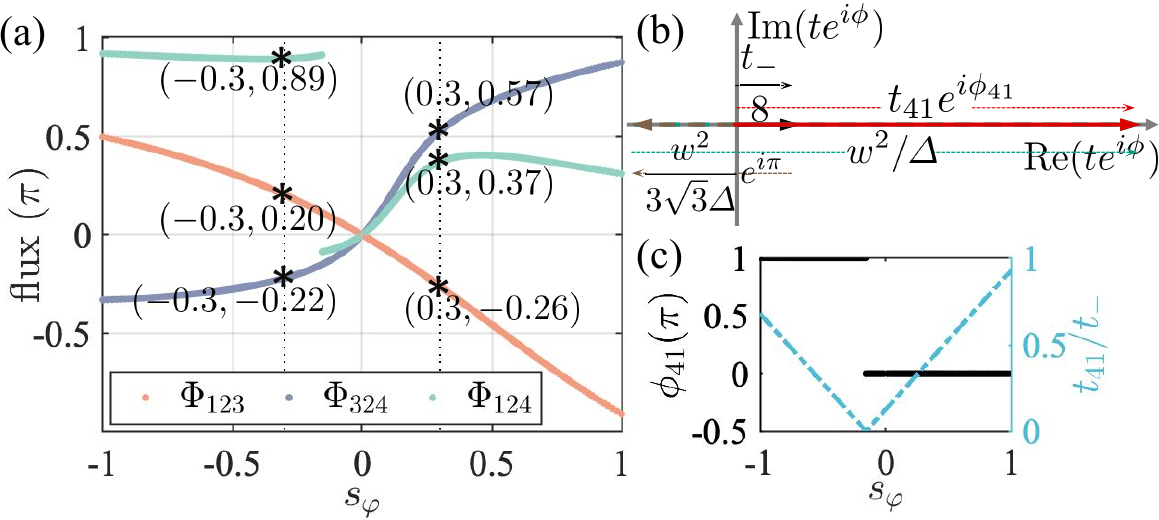}
\caption{Manipulating the synthetic flux.
(a) Variations of the synthetic fluxes under the control of $s_{\varphi}$. Here we take a vanishing decay rate $\gamma=0$. The asterisks indicate the values of synthetic fluxes for calculations in
Fig.~\ref{fig:3}.
(b) An exemplary illustration for constructing the hopping amplitude (up to second-order processes) from atom $4$ to $1$, where all components lie on the real axis.
(c) Variation of hopping amplitude from site $4$ to $1$ under the control of $s_{\varphi}$. When $t_{41}$ crosses zero, the phase $\phi_{41}$ undergoes a discontinuous jump from $\pi$ to $0$, leading to the discontinuity observed in $\Phi_{124}$, as shown in (a).}
\label{fig:2}
\end{figure}

In Fig.~\ref{fig:2}(a), we show the variation of the synthetic fluxes as functions of a dimensionless control parameter $s_{\varphi}=\tilde{\varDelta}^{2}/\tilde{w}$, where $\tilde{\varDelta}:=\varDelta/t_{-}$ and $\tilde{w}:=w/t_{-}$ (See Appendix for the definition and detailed derivation of $t_{-}$ and $w$). The discontinuity near $s_\varphi=-0.15$ in $\Phi_{124}$ originates from a vanishing $t_{41}$, due to the destructive interference of different processes [see Figs.~\ref{fig:2}(b)(c)]. Note that while we take $\gamma=0$ in Fig.~\ref{fig:2}, the dependence of the fluxes on $s_{\varphi}$ does note change dramatically under moderate $\gamma$.
In a simple triangular configuration, the sign of the flux determines the direction of chiral motion of the excitation. Likewise, in our setup, multiple synthetic fluxes compete with one another to influence the chiral motion of the injected excitation. As we illustrate below, the dominant impact comes from the competition between $\Phi_{123}$ and $\Phi_{324}$, because of the relatively larger hopping rates involved.


\begin{figure}[tbp]
\includegraphics[width=0.48\textwidth]{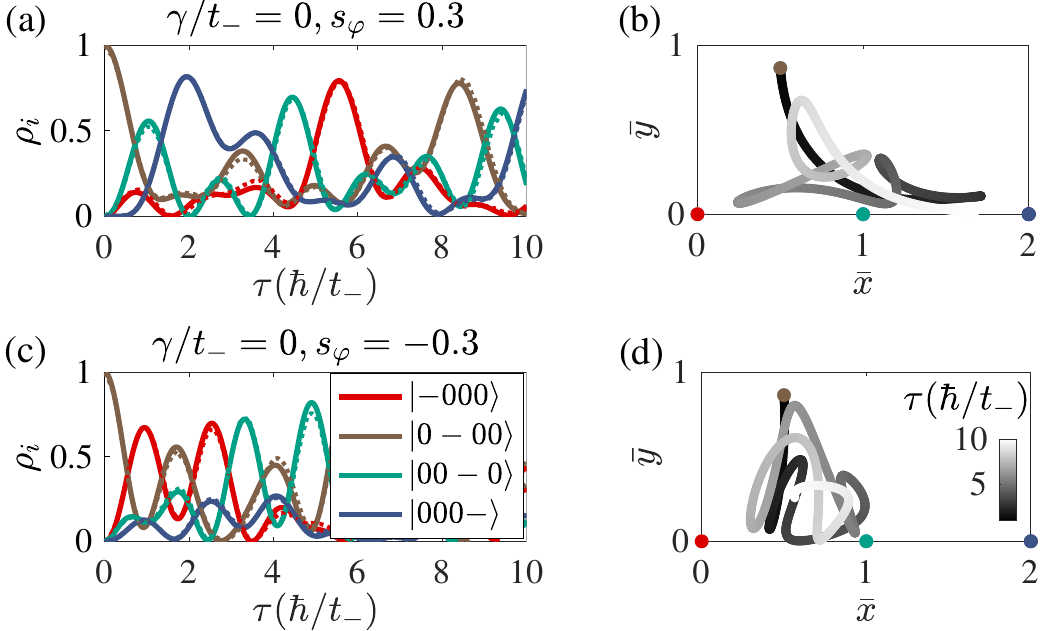}
\caption{The transport dynamics of a single excitation in the absence of dissipation.
(a) Time evolution of the occupation probabilities of states $\left|-000\right\rangle$, $\left|0-00\right\rangle$, $\left|00- 0\right\rangle$, and $\left|000-\right\rangle$, with $s_{\varphi}=0.3$. Solid lines represent simulation results under the effective Hamiltonian, while dashed lines correspond to results under the full Hamiltonian (see Appendix), which additionally includes the population of all relevant $\left|+\right\rangle$ excitations.
(b) The corresponding normalized trajectory of the excitation for the evolution in (a), where the change in color from dark to light represents the passage of time.
(c) Time evolution of the probabilities with $s_{\varphi}=-0.3$.
(d) The corresponding normalized trajectory of the excitation for the evolution in (c).
All relevant synthetic fluxes are indicated by asterisks in Fig.~\ref{fig:2}(a), and we take $\hbar/t_{-}$ as the unit of time.
}
\label{fig:3}
\end{figure}

\begin{figure}[tbp]
\includegraphics[width=0.48\textwidth]{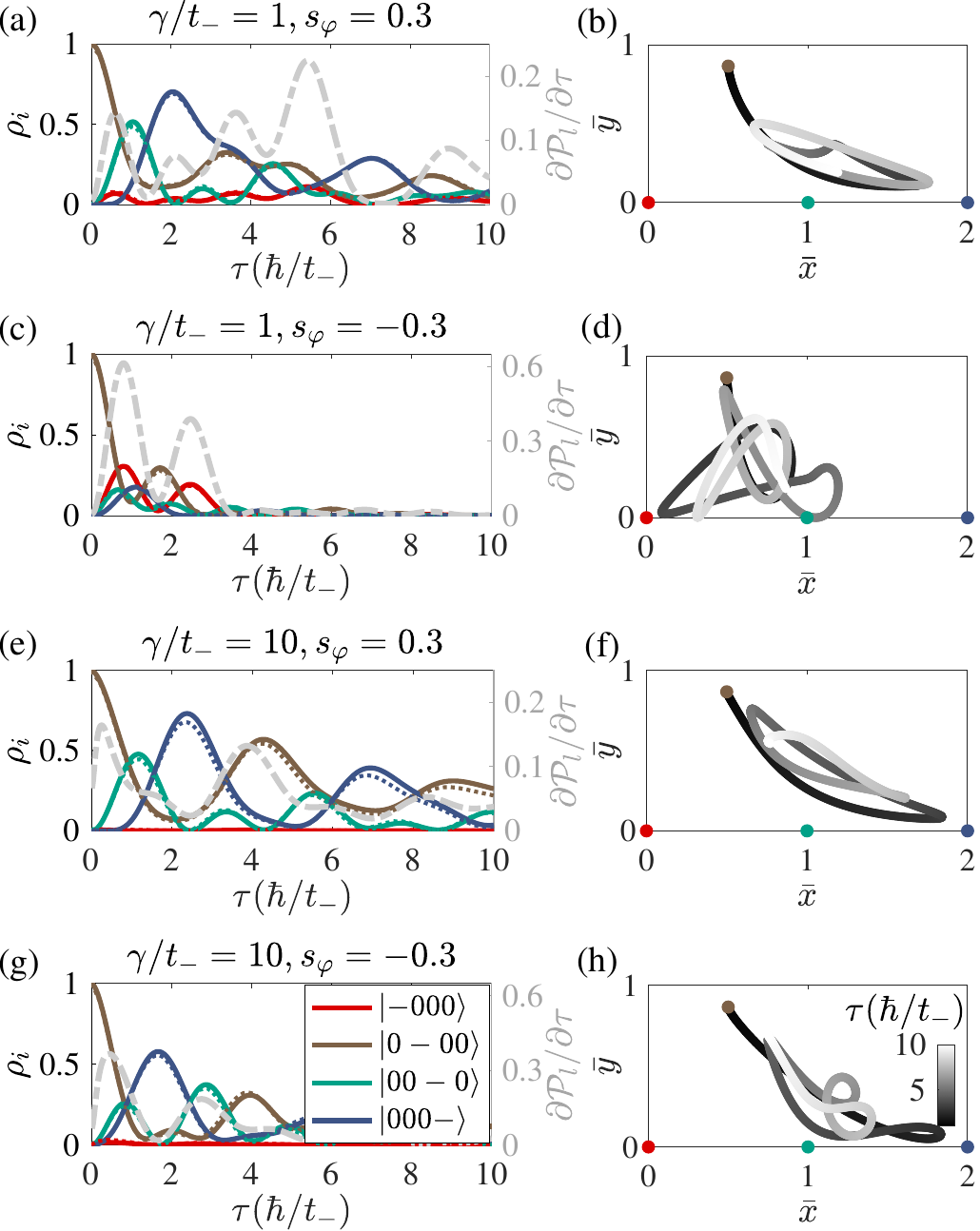}
\caption{The transport dynamics of a single excitation in the presence of dissipation.
(a) Time evolution of occupation probabilities, for $\gamma/t_-=1$ and $s_{\varphi}=0.3$. Solid lines represent simulation results under the effective Hamiltonian, while the short-dashed lines correspond to results under the full Hamiltonian.
(b) The corresponding normalized trajectory of the excitation for the evolution in (a).
(c) Time evolution of occupation probabilities, for $\gamma/t_-=1$ and $s_{\varphi}=-0.3$.
(d) The corresponding normalized trajectory of the excitation for the evolution in (c).
(e)(g) Time evolution of occupation probabilities, for $\gamma/t_-=10$ and $s_{\varphi}=\pm 0.3$.
(f)(h) The corresponding normalized trajectories of the excitation for the evolution in (e) and (g), respectively.
In (a)(c)(e)(g), the gray long-dashed lines represent the population loss rate (see main text for definition), with values referenced to the right $y$ axis.
}
\label{fig:4}
\end{figure}

\begin{figure}[tbp]
\includegraphics[width=0.48\textwidth]{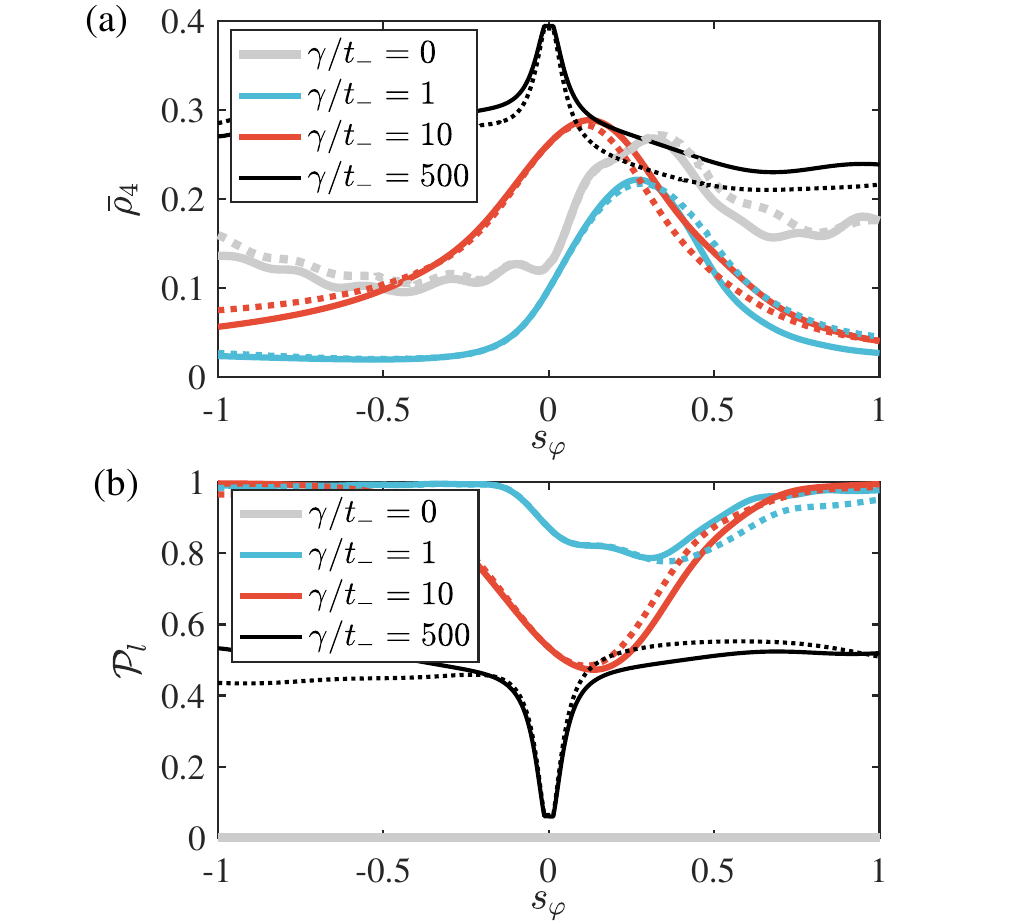}
\caption{
(a) Time-averaged population of the output site as a function of the control parameter $s_{\varphi}$, with a total evolution time $10\hbar/t_{-}$.
(b) The total population loss during an evolution period of $10\hbar/t_{-}$.
In both panels, the colored lines represent different decay rate, the solid lines are results calculated from the effective Hamiltonian $H_{\text{eff}}$, and the dashed lines correspond to those from the full Hamiltonian.
}
\label{fig:5}
\end{figure}

\begin{figure}[tbp]
\includegraphics[width=0.48\textwidth]{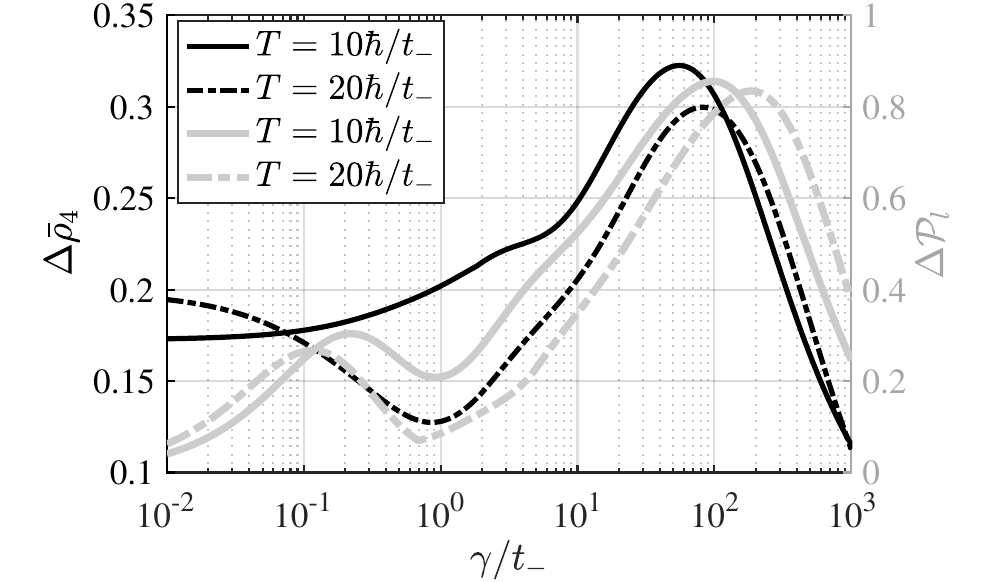}
\caption{Regime of tunable transport. Calculated $\bar{\rho}_{4}$ and $\mathcal{P}_{l}$ (see main text for definition) for different evolution time $\tau$ and
decay rate $\gamma$. The black solid (dashed) line is the calculated $\bar{\rho}_{4}$ for evolution time $\tau=10\hbar/t_-$ ($\tau=20\hbar/t_-$).
The gray solid (dashed) line is $\mathcal{P}_{l}$ for evolution time $\tau=10\hbar/t_-$ ($\tau=20\hbar/t_-$).
}
\label{fig:6}
\end{figure}

\section{Dissipationless dynamics under the synthetic fluxes}

We first study the transport of the injected excitation in the absence of dissipation, with $\gamma=0$. In Figs.~\ref{fig:3}(a)(c), we show the time-evolved probabilities $\rho_{i}$ for the excitation to occupy the $i$th site. Correspondingly, in Figs.~\ref{fig:3}(b)(d), we show the normalized trajectory of the excitation, which is defined as
\begin{align}
\bar{x}&=\frac{\sum_{i=1}^{4} x_{i}\rho_i}{r_0\sum_{i=1}^{4}\rho_{i}},\\
\bar{y}&=\frac{\sum_{i=1}^{4} y_{i}\rho_i}{r_0\sum_{i=1}^{4}\rho_{i}},
\end{align}
where $\left(x_i,\ y_i\right)$ is the coordinates of site $i$ (in units of $r_0$) in Figs.~\ref{fig:3}(b)(d),
$\tau$ represents the time of evolution, and $\left(\bar{x},\ \bar{y}\right)$ is the dimensionless coordinates of the normalized trajectory.

Notably, in Fig.~\ref{fig:3}(d), the trajectory is mostly within the triangle formed by sites $123$ (denoted as triangle $\Delta_{123}$), meaning the dominance of the occupation in these sites at all times.
This is in contrast to Fig.~\ref{fig:3}(b), where the trajectory is dominantly within the triangle $\Delta_{234}$, though the occupation of site $1$ can be dominant at intermediate times.
In both cases, the trajectories exhibit a CCW tendency.

Under the parameters of Figs.~\ref{fig:3}(a)(b), $\Phi_{123}<0$, $\Phi_{324}>0$, and $\Phi_{124}>0$.
Notice that $\Phi_{ijk}>0$ ($\Phi_{ijk}<0$) should lead to CCW (CW) chiral motion within the triangle $\Delta_{ijk}$, with the strongest CCW or CW tendency under $\Phi_{ijk}=\pm \pi/2$~\cite{Lienhard-20,Roushan-17} .
We therefore attribute the transport phenomena to the competition between $\Phi_{123}$ and $\Phi_{324}$.
For instance, in Figs.~\ref{fig:3}(a)(b), the excitation motion tends to be CW within the triangle $\Delta_{123}$, and CCW within the triangle $\Delta_{324}$. As a result, the excitation injected from site $2$ would predominantly flow to site $3$, and subsequently to site $4$, since $|\Phi_{324}|$ is closer to $\pi/2$ than $|\Phi_{123}|$. This is further facilitated by $\Phi_{124}>0$, which also favors a CCW transport.

On the other hand, in Figs.~\ref{fig:3}(c)(d), $\Phi_{123}>0$, $\Phi_{324}<0$, and $\Phi_{124}>0$.
The excitation motion tends to be CCW within the triangle $\Delta_{123}$, and CW within the triangle $\Delta_{324}$. Since the hopping amplitude between sites $2$ and $4$ is smaller than that between $2$ and $1$, and since $\Phi_{124}>0$, the excitation injected from site $2$ would predominantly flow to site $1$. Further, once the excitation moves to site $3$, both $\Phi_{123}$ and $\Phi_{324}$ favor the subsequent transport to site $2$. These factors lead to a CCW transport with very small occupation of site $4$.

\section{Dynamics with dissipation}

We now switch on the on-site light-induced dissipation at site $1$. When the decay rate $\gamma$ is small, the trajectories remain qualitatively similar to the case of $\gamma=0$, as shown in Figs.~\ref{fig:4}(a)(c). However, the two trajectories are now predominantly limited within the triangles $\Delta_{324}$ (for $s_{\varphi}=0.3$) and $\Delta_{123}$ (for $s_{\varphi}=-0.3$), respectively.
This gives rise to a considerable difference in the time-averaged population of site $4$, as we will explicitly demonstrate later. The difference between the trajectories in Fig.~\ref{fig:3}(b) and Fig.~\ref{fig:4}(b) arises from the interplay of dissipation and synthetic flux.
To further illustrate this point, we define the total population loss as $\mathcal{P}_{l}=1-\sum_{i=1}^{4}\rho_{i}$, as well as the loss rate  $\partial\mathcal{P}_{l}/\partial \tau$.
In Figs.~\ref{fig:4}(a)(c), peaks of the loss rate  $\partial\mathcal{P}_{l}/\partial \tau$ almost overlaps with the occupation of site $1$, suggesting a rapid loss from site $1$.


In the case of Figs.~\ref{fig:4}(a)(b), the dissipation depletes the occupation in site $1$, which limits the trajectory to the triangle $\Delta_{324}$, giving rise to a large time-averaged population of site $4$.
By contrast, in the case of Figs.~\ref{fig:4}(c)(d), the appreciable population of site $1$ at short times under the CCW chiral motion leads to significant loss. The time-averaged transfer to site $4$ becomes suppressed.

When the decay rate is further increased, the loss is so significant that the occupation in site $1$ is completely suppressed. Notice that this already occurs when the system is still far from the quantum Zeno regime (see discussions in the next section). This is the case for $\gamma/t_-=10$, as shown in Figs.~\ref{fig:4}(e)(g),
where the vanishing population of site $1$ is accompanied by large loss (long dashed). The trajectories on the other hand, become limited to the triangle $\Delta_{324}$. Importantly, throughout this intermediate-$\gamma$ regime, the time-averaged transport to site $4$ can be switched on and off by tuning the flux parameter $s_{\varphi}$.

\section{Further analysis of loss}

To provide further evidence for the analysis above, in Fig.~\ref{fig:5}, we show the time-averaged population of the output site under different decay rate, for which we define
\begin{align}
\bar{\rho}_{4}=\frac{1}{T}\int_{0}^{T}d\tau\rho_{4}(\tau),
\end{align}
which is a function of the control parameter $s_{\varphi}$.
As shown  in Fig.~\ref{fig:5}(a), in the presence of a finite $\gamma$ (but not large enough to reach the quantum Zeno regime), the transport to site $4$ remains small but for an intermediate regime in $s_{\varphi}$. Such a regime also features suppressed loss, as illustrated in Fig.~\ref{fig:5}(b). By contrast, in the regime where the transport to site $4$ is suppressed, the loss becomes significant.
Hence, combining dissipation and synthetic flux, we achieve an efficient control over the transport of injected excitation from site $2$ to the output site $4$.

Finally, in Fig.~\ref{fig:6}, we show the range of variation of $\bar{\rho}_{4}$ and $\mathcal{P}_{l}$ for $s_{\varphi}\in[-1,1]$. For this purpose, we define
$\Delta\bar{\rho}_{4}=\max\left(\bar{\rho}_{4}\right)-\min\left(\bar{\rho}_{4}\right)$ and $\Delta\mathcal{P}_{l}=\max\left(\mathcal{P}_{l}\right)-\min\left(\mathcal{P}_{l}\right)$,
where the function $\max$ ($\min$) gives the maximum (minimum) value of the corresponding variable when $s_{\varphi}$ varies in the range of $[-1,1]$.

A regime of tunable transport to site $4$ is indicated by a large $\Delta\bar{\rho}_{4}$. The overlap of the peaks of $\Delta\bar{\rho}_{4}$ and $\Delta\mathcal{P}_{l}$ suggests that the tunable transport is indispensable with large variations in the dissipation. In the large-$\gamma$ limit, the transport dynamics become restricted to the triangle $\Delta_{324}$, as both the occupation of site $1$ and the total loss are significantly suppressed due to the quantum Zeno effect~\cite{Itano-90}. The tunable transport is no longer achievable in this regime.

\section{Conclusion}
We show that the interplay of interaction-induced synthetic flux and dissipation gives rise to tunable excitation dynamics in Rydberg atoms. Considering a four-site configuration as a concrete example, we demonstrate how the injected single excitation can be transported to an output site, by tuning the flux and dissipation. This is further confirmed by characterizing the total loss rate and
probability evolution throughout the transport dynamics.
While the light-induced dissipation we consider here gives rise to non-Hermiticity through post selection, it is also interesting to study the transport dynamics in the context of quantum open systems, where the density-matrix dynamics of the system is driven by Liouvillian superoperators~\cite{Petrosyan-13,Rao-14}. Another interesting scenario is the dynamics of several excitations, where the interaction-induced gauge field become density-dependent and acquires its own dynamics in the context of quantum many-body open systems.

\acknowledgments{This research is supported by the Natural Science Foundation of China (Grant No. 11974331).

\begin{widetext}

		\renewcommand{\thesection}{\Alph{section}}
		\renewcommand{\thefigure}{A\arabic{figure}}
		\renewcommand{\thetable}{A\Roman{table}}
		\setcounter{figure}{0}
		\renewcommand{\theequation}{A\arabic{equation}}
		\setcounter{equation}{0}
    \appendix*

    \section{Derivation of the effective non-Hermitian Hamiltonian}

In this Appendix, we derive the effective Hamiltonian from the full Hamiltonian where both $|\pm\rangle$ excitations are considered. According to Fig.~\ref{fig:1},
in the subspace spanned by the states $\left\{ \left|\pm 000\right\rangle ,\left|0\pm 00\right\rangle , \left|00\pm 0\right\rangle ,\left|000\pm\right\rangle \right\} $, the system is described by the following full Hamiltonian
\begin{equation}
H_{\text{full}}=\left(\begin{array}{cccccccc}
\frac{\varDelta}{2}-i\gamma & t_{-} & t_{-} & \frac{t_{-}}{8} & 0 & we^{2i\varphi_{12}} & we^{2i\varphi_{13}} & \frac{we^{2i\varphi_{14}}}{8}\\
t_{-} & \frac{\varDelta}{2} & t_{-} & \frac{t_{-}}{3\sqrt{3}} & we^{2i\varphi_{21}} & 0 & we^{2i\varphi_{23}} & \frac{we^{2i\varphi_{24}}}{3\sqrt{3}}\\
t_{-} & t_{-} & \frac{\varDelta}{2} & t_{-} & we^{2i\varphi_{31}} & we^{2i\varphi_{32}} & 0 & we^{2i\varphi_{34}}\\
\frac{t_{-}}{8} & \frac{t_{-}}{3\sqrt{3}} & t_{-} & \frac{\varDelta}{2} & \frac{we^{2i\varphi_{41}}}{8} & \frac{we^{2i\varphi_{42}}}{3\sqrt{3}} & we^{2i\varphi_{43}} & 0\\
0 & we^{-2i\varphi_{12}} & we^{-2i\varphi_{13}} & \frac{we^{-2i\varphi_{14}}}{8} & -\frac{\varDelta}{2}-i\gamma & t_{+} & t_{+} & \frac{t_{+}}{8}\\
we^{-2i\varphi_{21}} & 0 & we^{-2i\varphi_{23}} & \frac{we^{-2i\varphi_{24}}}{3\sqrt{3}} & t_{+} & -\frac{\varDelta}{2} & t_{+} & \frac{t_{+}}{3\sqrt{3}}\\
we^{-2i\varphi_{31}} & we^{-2i\varphi_{32}} & 0 & we^{-2i\varphi_{34}} & t_{+} & t_{+} & -\frac{\varDelta}{2} & t_{+}\\
\frac{we^{-2i\varphi_{41}}}{8} & \frac{we^{-2i\varphi_{42}}}{3\sqrt{3}} & we^{-2i\varphi_{43}} & 0 & \frac{t_{+}}{8} & \frac{t_{+}}{3\sqrt{3}} & t_{+} & -\frac{\varDelta}{2}
\end{array}\right).
\end{equation}
The parameter $\varDelta$ is the energy difference between
the state $\left|+\right\rangle $ and the state $\left|-\right\rangle $. The hopping amplitudes are given by
\begin{align}
t_{\pm}&=\frac{\left|\left\langle \pm\right|\hat{d}^{+}\left|0\right\rangle \right|^{2}}{8\pi\epsilon_{0}r_{0}^{3}}\\
w&=-\frac{3\left\langle +\right|\hat{d}^{+}\left|0\right\rangle \left\langle 0\right|\hat{d}^{+}\left|-\right\rangle }{8\pi\epsilon_{0}r_{0}^{3}}.
\end{align}
Additionally, the van der Waals interaction is neglected in the parameter regime we consider, since
it is much weaker compared to the resonant dipole-dipole interaction.

By defining the creation and annihilation operators on site $i$,
$a_{i}^{\dagger}\left|0\right\rangle =\left|-\right\rangle _{i}$
and $b_{i}^{\dagger}\left|0\right\rangle =\left|+\right\rangle _{i}$,
the Hamiltonian is expressed as
\begin{equation}
H_{\text{full}}=-i\gamma\left(a_{1}^{\dagger}a_{1}+b_{1}^{\dagger}b_{1}\right)+\frac{\varDelta}{2}\underset{i}{\sum}\left(a_{i}^{\dagger}a_{i}-b_{i}^{\dagger}b_{i}\right)+\underset{i\neq j}{\sum}\left(\frac{r_{0}}{r_{ij}}\right)^{3}\left(t_{-}a_{i}^{\dagger}a_{j}+t_{+}b_{i}^{\dagger}b_{j}+we^{2i\varphi_{ij}}a_{i}^{\dagger}b_{j}+we^{-2i\varphi_{ij}}b_{i}^{\dagger}a_{j}\right).
\end{equation}
We work in the regime with $\varDelta\gg t_{\pm},w$, so that when
we take $\left|0-00\right\rangle $ as the initial state, the system primarily remains in the subspace spanned by the states $\left\{ \left|-000\right\rangle ,\left|0-00\right\rangle ,\left|00-0\right\rangle ,\left|000-\right\rangle \right\} $.
Adiabatically eliminating the $|+\rangle$ excitations thus lead to the effective Hamiltonian
\begin{equation}
H_{\text{eff}}=\sum_{i}\varepsilon_{i}a_{i}^{\dagger}a_{i}+\sum_{ij}t_{ij}e^{i\phi_{ij}}a_{i}^{\dagger}a_{j},
\end{equation}
where $\varepsilon_{i}$ are the self-energy terms
\begin{align}
\varepsilon_{1} & =-i\gamma+\frac{\varDelta}{2}+\frac{w^{2}}{\varDelta}\left[e^{2i\left(\varphi_{12}-\varphi_{21}\right)}+e^{2i\left(\varphi_{13}-\varphi_{31}\right)}+\frac{1}{64}e^{2i\left(\varphi_{14}-\varphi_{41}\right)}\right]\nonumber\\
 & =-i\gamma+\frac{\varDelta}{2}+\frac{w^{2}}{\varDelta}\frac{129}{64},\\
\varepsilon_{2} & =\frac{\varDelta}{2}+\frac{w^{2}}{\varDelta}\left[\frac{\varDelta}{\varDelta+i\gamma}e^{2i\left(\varphi_{21}-\varphi_{12}\right)}+e^{2i\left(\varphi_{23}-\varphi_{32}\right)}+\frac{1}{27}e^{2i\left(\varphi_{24}-\varphi_{42}\right)}\right]\nonumber\\
 & =\frac{\varDelta}{2}+\frac{w^{2}}{\varDelta}\left(\frac{\varDelta^{2}-i\varDelta\gamma}{\varDelta^{2}+\gamma^{2}}+\frac{28}{27}\right),\\
\varepsilon_{3} & =\frac{\varDelta}{2}+\frac{w^{2}}{\varDelta}\left[\frac{\varDelta}{\varDelta+i\gamma}e^{2i\left(\varphi_{31}-\varphi_{13}\right)}+e^{2i\left(\varphi_{32}-\varphi_{23}\right)}+e^{2i\left(\varphi_{34}-\varphi_{43}\right)}\right]\nonumber\\
 & =\frac{\varDelta}{2}+\frac{w^{2}}{\varDelta}\left(\frac{\varDelta^{2}-i\varDelta\gamma}{\varDelta^{2}+\gamma^{2}}+2\right),\\
\varepsilon_{4} & =\frac{\varDelta}{2}+\frac{w^{2}}{\varDelta}\left[\frac{\varDelta}{\varDelta+i\gamma}\frac{1}{64}e^{2i\left(\varphi_{41}-\varphi_{14}\right)}+\frac{1}{27}e^{2i\left(\varphi_{42}-\varphi_{24}\right)}+e^{2i\left(\varphi_{43}-\varphi_{34}\right)}\right]\nonumber\\
 & =\frac{\varDelta}{2}+\frac{w^{2}}{\varDelta}\left(\frac{\varDelta^{2}-i\varDelta\gamma}{\varDelta^{2}+\gamma^{2}}\frac{1}{64}+\frac{28}{27}\right),
\end{align}
and $t_{ij}e^{i\phi_{ij}}$ are the hopping terms, which satisfy $t_{ji}=t_{ij}$ and $\phi_{ji}=-\phi_{ij}$, with
\begin{align}
t_{12}e^{i\phi_{12}} & =t_{-}+\frac{w^{2}}{\varDelta}\left(e^{2i\left(\varphi_{23}-\varphi_{31}\right)}+\frac{1}{24\sqrt{3}}e^{2i\left(\varphi_{24}-\varphi_{41}\right)}\right)\nonumber\\
&=t_{-}+\frac{w^{2}}{\varDelta}\left[\left(-\frac{1}{2}-i\frac{\sqrt{3}}{2}\right)+\frac{1}{24\sqrt{3}}\left(\frac{1}{2}-i\frac{\sqrt{3}}{2}\right)\right],\\
t_{13}e^{i\phi_{13}} & =t_{-}+\frac{w^{2}}{\varDelta}\left(e^{2i\left(\varphi_{32}-\varphi_{21}\right)}+\frac{1}{8}e^{2i\left(\varphi_{34}-\varphi_{41}\right)}\right)\nonumber\\
&=t_{-}+\frac{w^{2}}{\varDelta}\left[\left(-\frac{1}{2}+i\frac{\sqrt{3}}{2}\right)+\frac{1}{8}\right],\\
t_{14}e^{i\phi_{14}} & =\frac{t_{-}}{8}+\frac{w^{2}}{\varDelta}\left(\frac{1}{3\sqrt{3}}e^{2i\left(\varphi_{42}-\varphi_{21}\right)}+e^{2i\left(\varphi_{43}-\varphi_{31}\right)}\right)\nonumber\\
&=\frac{t_{-}}{8}+\frac{w^{2}}{\varDelta}\left(-\frac{1}{3\sqrt{3}}+1\right),\\
t_{23}e^{i\phi_{23}} & =t_{-}+\frac{w^{2}}{\varDelta}\left(\frac{\varDelta}{\varDelta+i\gamma}e^{2i\left(\varphi_{31}-\varphi_{12}\right)}+\frac{1}{3\sqrt{3}}e^{2i\left(\varphi_{34}-\varphi_{42}\right)}\right)\nonumber\\
&=t_{-}+\frac{w^{2}}{\varDelta}\left[\frac{\varDelta^{2}-i\varDelta\gamma}{\varDelta^{2}+\gamma^{2}}\left(-\frac{1}{2}-i\frac{\sqrt{3}}{2}\right)+\frac{1}{3\sqrt{3}}\left(\frac{1}{2}+i\frac{\sqrt{3}}{2}\right)\right],\\
t_{24}e^{i\phi_{24}} & =\frac{t_{-}}{3\sqrt{3}}+\frac{w^{2}}{\varDelta}\left(\frac{1}{8}\frac{\varDelta}{\varDelta+i\gamma}e^{2i\left(\varphi_{41}-\varphi_{12}\right)}+e^{2i\left(\varphi_{43}-\varphi_{32}\right)}\right)\nonumber\\
&=\frac{t_{-}}{3\sqrt{3}}+\frac{w^{2}}{\varDelta}\left[\frac{1}{8}\frac{\varDelta^{2}-i\varDelta\gamma}{\varDelta^{2}+\gamma^{2}}\left(-\frac{1}{2}-i\frac{\sqrt{3}}{2}\right)+\left(-\frac{1}{2}+i\frac{\sqrt{3}}{2}\right)\right],\\
t_{34}e^{i\phi_{34}} & =t_{-}+\frac{w^{2}}{\varDelta}\left(\frac{1}{8}\frac{\varDelta}{\varDelta+i\gamma}e^{2i\left(\varphi_{41}-\varphi_{13}\right)}+\frac{1}{3\sqrt{3}}e^{2i\left(\varphi_{42}-\varphi_{23}\right)}\right)\nonumber\\
&=t_{-}+\frac{w^{2}}{\varDelta}\left[\frac{1}{8}\frac{\varDelta^{2}-i\varDelta\gamma}{\varDelta^{2}+\gamma^{2}}+\frac{1}{3\sqrt{3}}\left(\frac{1}{2}+i\frac{\sqrt{3}}{2}\right)\right].
\end{align}



\begin{thebibliography}{99}
\bibitem{reviewRdB-1}
M. Saffman, T. G. Walker, and K. Mølmer, Rev. Mod. Phys. {\bf 82}, 2313 (2010).
\bibitem{reviewRdB-2}
M. Morgado and S. Whitlock, AVS Quantum Science {\bf 3}, 023501 (2021).
\bibitem{reviewRdB-3}
X. Wu, X. Liang, Y. Tian, F. Yang, C. Chen, Y.-C. Liu, M. K. Tey, and L. You, Chin. Phys. B {\bf 30}, 020305 (2021).
\bibitem{reviewRdB-4}
 A. Browaeys and T. Lahaye, Nat. Phys. {\bf 16}, 132 (2020).

\bibitem{Ashkin-86}
A. Ashkin, J. M. Dziedzic, J. E. Bjorkholm, and S. Chu,
Opt. Lett. {\bf 11}, 288 (1986).
\bibitem{Ebadi-21}
S. Ebadi, T. T. Wang, H. Levine, A. Keesling, G. Semeghini, A. Omran, D. Bluvstein, R. Samajdar, H. Pichler, W. W. Ho, S. Choi, S. Sachdev, M. Greiner, V. Vuleti\'{c}, and M. D. Lukin, Nature {\bf 595}, 227 (2021).
\bibitem{Morgado-21}
M. Morgado and S. Whitlock, AVS Quantum Science {\bf 3}, 023501 (2021).
\bibitem{Scholl-21}
P. Scholl, M. Schuler, H. J. Williams, A. A. Eberharter, D. Barredo, K.-N. Schymik, V. Lienhard, L.-P. Henry, T. C. Lang, T. Lahaye, A. M. L{\"{a}}uchli, and A. Browaeys, Nature {\bf 595}, 233 (2021).

\bibitem{Weber-17}
S. Weber, C. Tresp, H. Menke, A. Urvoy, O. Firstenberg, H. P. B{\"{u}}chler, and S. Hofferberth, J. Phys. B {\bf 50}, 133001 (2017).
\bibitem{Browaeys-16}
A. Browaeys, D. Barredo, and T. Lahaye, J. Phys. B {\bf 49}, 152001 (2016).

\bibitem{Leseleuc-18}
S. de L\'{e}s\'{e}leuc, S. Weber, V. Lienhard, D. Barredo, H. P. B{\"{u}}chler, T. Lahaye, and A. Browaeys,
Phys. Rev. Lett. {\bf 120}, 113602 (2018).
\bibitem{Bernien-17}
H. Bernien, S. Schwartz, A. Keesling, H. Levine, A. Omran, H. Pichler, S. Choi, A. S. Zibrov, M. Endres, M. Greiner, V. Vuleti\'{c}, and M. D. Lukin, Nature {\bf 551}, 579 (2017).
\bibitem{Schauss-18}
P. Schauss, Quantum Sci. Technol. {\bf 3}, 023001 (2018).
\bibitem{Barredo-15}
D. Barredo, H. Labuhn, S. Ravets, T. Lahaye, A. Browaeys, and C. S. Adams,
Phys. Rev. Lett. {\bf 114}, 113002 (2015).
\bibitem{Signoles-21}
A. Signoles, T. Franz, R. Ferracini Alves, M. G{\"{a}}rttner, S. Whitlock, G. Zürn, and M. Weidemüller,
Phys. Rev. X {\bf 11}, 011011 (2021).
\bibitem{Steinert-23}
L.-M. Steinert, P. Osterholz, R. Eberhard, L. Festa, N. Lorenz, Z. Chen, A. Trautmann, and C. Gross, Phys. Rev. Lett. {\bf 130}, 243001 (2023).

\bibitem{Leseleuc-19}
S. de L\'{e}s\'{e}leuc, V. Lienhard, P. Scholl, D. Barredo, S. Weber, N. Lang, H. P. B{\"{u}}chler, T. Lahaye, and A. Browaeys, Science {\bf 365}, 775 (2019).
\bibitem{Verresen-21}
R. Verresen, M. D. Lukin, and A. Vishwanath, Phys. Rev. X {\bf 11}, 031005 (2021).
\bibitem{Weber-22}
S. Weber, R. Bai, N. Makki, J. M{\"{o}}gerle, T. Lahaye, A. Browaeys, M. Daghofer, N. Lang, and H. P. B{\"{u}}chler, PRX Quantum {\bf 3}, 030302 (2022).

\bibitem{Zhang-18}
J. Zhang, J. Unmuth-Yockey, J. Zeiher, A. Bazavov, S.-W. Tsai, and Y. Meurice, Phys. Rev. Lett. {\bf 121}, 223201 (2018).
\bibitem{Celi-20}
A. Celi, B. Vermersch, O. Viyuela, H. Pichler, M. D. Lukin, and P. Zoller, Phys. Rev. X {\bf 10}, 021057 (2020).
\bibitem{Notarnicola-20}
S. Notarnicola, M. Collura, and S. Montangero, Phys. Rev. Res. {\bf 2}, 013288 (2020).
\bibitem{Surace-20}
F. M. Surace, P. P. Mazza, G. Giudici, A. Lerose, A. Gambassi, and M. Dalmonte, Phys. Rev. X {\bf 10}, 021041 (2020).
\bibitem{Cuadra-22}
D. Gonz\'{a}lez-Cuadra, T. V. Zache, J. Carrasco, B. Kraus, and P. Zoller, Phys. Rev. Lett. {\bf 129}, 160501 (2022).

\bibitem{Giudici-22}
G. Giudici, M. D. Lukin, and H. Pichler, Phys. Rev. Lett. {\bf 129}, 090401 (2022).
\bibitem{Semeghini-21}
G. Semeghini, H. Levine, A. Keesling, S. Ebadi, T. T. Wang, D. Bluvstein, R. Verresen, H. Pichler, M. Kalinowski, R. Samajdar, A. Omran, S. Sachdev, A. Vishwanath, M. Greiner, V. Vuleti\'{c}, and M. D. Lukin, Science {\bf 374}, 1242 (2021).
\bibitem{Cheng-23}
Y. Cheng, C. Li, and H. Zhai, New J. Phys. {\bf 25}, 033010 (2023).
\bibitem{Kalinowski-23}
M. Kalinowski, N. Maskara, and M. D. Lukin, Phys. Rev. X {\bf 13}, 031008 (2023).
\bibitem{Ohler-23}
S. Ohler, M. Kiefer-Emmanouilidis, and M. Fleischhauer, Phys. Rev. Res. {\bf 5}, 013157 (2023).
\bibitem{Lienhard-20}
V. Lienhard, P. Scholl, S. Weber, D. Barredo, S. de L\'{e}s\'{e}leuc, R. Bai, N. Lang, M. Fleischhauer, H. P. B{\"{u}}chler, T. Lahaye, and A. Browaeys, Phys. Rev. X {\bf 10}, 021031 (2020).

\bibitem{Zhao-23}
Y. Zhao and X.-F. Shi, arXiv:2206.04213.
\bibitem{F-23}
T.-F. J. Poon, X.-C. Zhou, B.-Z. Wang, T.-H. Yang, and X.-J. Liu, arXiv:2302.13104.
\bibitem{Yang-22}
T.-H. Yang, B.-Z. Wang, X.-C. Zhou, and X.-J. Liu, Phys. Rev. A {\bf 106}, L021101 (2022).


\bibitem{Wu-22}
X. Wu, F. Yang, S. Yang, K. Mølmer, T. Pohl, M. K. Tey, and L. You, Phys. Rev. Res. {\bf 4}, L032046 (2022).
\bibitem{Li-22}
X. X. Li, J. B. You, X. Q. Shao, and W. Li, Phys. Rev. A {\bf 105}, 032417 (2022).

\bibitem{reviewNH-1}
Y. Ashida, Z. Gong, and M. Ueda, Adv. Phys. {\bf 69}, 249 (2020).
\bibitem{Shen-23}
R. Shen, T. Chen, M. M. Aliyu, F. Qin, Y. Zhong, H. Loh, and C. H. Lee, Phys. Rev. Lett. {\bf 131}, 080403 (2023).

\bibitem{Peter-15}
D. Peter, N. Y. Yao, N. Lang, S. D. Huber, M. D. Lukin, and H. P. B{\"{u}}chler, Phys. Rev. A {\bf 91}, 053617 (2015).

\bibitem{Roushan-17}
P. Roushan, C. Neill, A. Megrant, Y. Chen, R. Babbush, R. Barends, B. Campbell, Z. Chen, B. Chiaro, A. Dunsworth, A. Fowler, E. Jeffrey, J. Kelly, E. Lucero, J. Mutus, P. J. J. O\'{M}alley, M. Neeley, C. Quintana, D. Sank, A. Vainsencher, J. Wenner, T. White, E. Kapit, H. Neven, and J. Martinis, Nat. Phys. {\bf 13}, 146 (2017).

\bibitem{Itano-90}
W. M. Itano, D. J. Heinzen, J. J. Bollinger, and D. J. Wineland, Phys. Rev. A {\bf 41}, 2295 (1990).

\bibitem{Petrosyan-13}
D. Petrosyan and K. M{\o}lmer, Phys. Rev. A {\bf 87}, 033416 (2013).
\bibitem{Rao-14}
D. D. B. Rao and K. M{\o}lmer, Phys. Rev. A {\bf 90}, 062319 (2014).

\end{thebibliography}

\end{widetext}
\bibliographystyle{apsrev4-1}

\end{document}